\documentclass{aa}
\usepackage{graphicx}
\usepackage{txfonts}
\usepackage{lscape}
\usepackage{natbib}
\bibpunct{(}{)}{;}{a}{}{,}

\newcommand{\ctss}{\mbox{counts s}^{-1}}
\newcommand{\fluxA}{\mbox{erg cm}^{-2}\ \mbox{s}^{-1}\ \mbox{\AA}^{-1}}
\newcommand{\flux}{\mbox{erg cm}^{-2}\ \mbox{s}^{-1}}
\newcommand{\lum}{d_{\mathrm{500 pc}}^2\ \mbox{erg s}^{-1}}
\newcommand{\lumrxj}{d_{\mathrm{600 pc}}^2\ \mbox{erg s}^{-1}}
\newcommand{\mdot}{d_{\mathrm{500 pc}}^2\ \mbox{g s}^{-1}}
\newcommand{\mdotrxj}{d_{\mathrm{600 pc}}^2\ \mbox{g s}^{-1}}
\newcommand{\mdotn}{\mbox{g s}^{-1}}
\newcommand{\magmom}{d_{\mathrm{500 pc}}\ \mbox{G cm}^3}

\begin{document}

\title{Broad-band properties of the hard X-ray cataclysmic variables IGR~J00234+6141 and 1RXS~J213344.1+510725
\thanks{Based on observations obtained with \emph{XMM-Newton} and \emph{INTEGRAL}, ESA science missions
with instruments and contributions directly funded by ESA Member States and NASA, and with \emph{Suzaku}, a
Japan's mission developed at the Institute of Space and Astronautical Science of Japan Aerospace Exploration
Agency in collaboration with U.S. (NASA/GSFC, MIT) and Japanese institutions.}}

\author{G.~Anzolin\inst{1,2}
\and
D.~de~Martino\inst{2}
\and
M.~Falanga\inst{3}
\and
K.~Mukai\inst{4}
\and
J.-M.~Bonnet-Bidaud\inst{3}
\and
M.~Mouchet\inst{5}
\and
Y.~Terada\inst{6}
\and
M.~Ishida\inst{7}
}

\institute{
Dipartimento di Astronomia, Universit\`{a} di Padova, vicolo dell'Osservatorio 3, I-35122 Padova, Italy\\
\email{gabriele.anzolin@unipd.it}
\and
INAF - Osservatorio Astronomico di Capodimonte, salita Moiariello 16, I-80131 Napoli, Italy \\
\email{demartino@oacn.inaf.it}
\and
CEA Saclay, DSM/DAPNIA/Service d'Astrophysique, F-91191 Gif-sur-Yvette, France \\
\email{[mfalanga;bonnetbidaud]@cea.fr}
\and
CRESST and X-Ray Astrophysics Laboratory, NASA Goddard Space Flight Center, Greenbelt, MD 20771,
USA and Department of Physics, University of Maryland, Baltimore County, 1000 Hilltop Circle,
Baltimore, MD 21250, USA \\
\email{koji.mukai@nasa.gov}
\and
Laboratoire APC, Universit\'{e} Denis Diderot, 10 rue Alice Domon et L\'{e}onie Duquet, F-75005
Paris, France and LUTH, Observatoire de Paris, Section de Meudon, 5 place Jules Janssen, F-92195
Meudon, France \\
\email{martine.mouchet@obspm.fr}
\and
Department of physics, Saitama University, Saitama, Japan \\
\email{terada@phy.saitama-u.ac.jp}
\and
Institute of Space and Astronautical Science, 3-1-1, Yoshinodai, Sagamihara, Kanagawa 229-8150, Japan \\
\email{ishida@astro.isas.jaxa.jp}
}

\date{Received ...; accepted ...}

\abstract
{A significant number of cataclysmic variables were detected as hard X-ray sources in the \emph{INTEGRAL} survey,
most of them of the magnetic intermediate polar type.}
{We present a detailed X-ray broad-band study of two new sources, IGR~J00234+6141 and 1RXS~J213344.1+510725,
that allow us to classify them as secure members of the intermediate polar class.}
{Timing and spectral analysis of IGR~J00234+6141 are based on a \emph{XMM-Newton} observation and \emph{INTEGRAL}
publicly available data. For 1RXS~J213344.1+510725 we use \emph{XMM-Newton} and \emph{Suzaku} observations at
different epochs, as well as \emph{INTEGRAL} publicly available data.} 
{We determine a spin period of $561.64 \pm 0.56$ s for the white dwarf in IGR~J00234+6141. The X-ray pulses are
observed up to $\sim 2$ keV. From \emph{XMM-Newton} and \emph{Suzaku} observations of 1RXS~J213344.1+510725,
we find a rotational period of $570.862 \pm 0.034$ s. The observations span three epochs where the pulsation
is observed to change at different energies both in amplitude and shape. In both objects, the spectral analysis
spanned over a wide energy range, from 0.3 to 100 keV, shows the presence of multiple emission components
absorbed by dense material. The X-ray spectrum of IGR~J00234+6141 is consistent with a multi-temperature plasma
with a maximum temperature of $\sim 50$ keV. In 1RXS~J213344.1+510725, multiple optically thin components are
inferred, as well as an optically thick (blackbody) soft X-ray emission with a temperature of $\sim 100$ eV.
This latter adds 1RXS~J213344.1+510725 to the growing group of soft X-ray intermediate polars.}
{These two intermediate polars, though showing similar rotational periods and being hard X-ray sources, appear
different in several respects, among which the presence of a soft X-ray component in 1RXS~J213344.1+510725.
This source also emits circularly polarized light in the optical band, thus joining as a fifth member the group
of soft and polarized intermediate polars. How the hard X-ray and polarized emission are thermalized in these
systems is a timely question.}


\titlerunning{Broad-band properties of the hard X-ray CVs IGR~J00234+6141 and 1RXS~J213344.1+510725}

\maketitle

\section{Introduction}

The first deep survey above 20 keV performed by the \emph{INTEGRAL} satellite has allowed the detection
of more than 400 X-ray sources \citep{bird07}. The extensive survey of the Galactic plane has revealed
that the contribution of Galactic X-ray binaries, especially the cataclysmic variable (CV) type, is
non-negligible in hard X-rays  \citep{sazonov06}. Most of the hard X-ray emitting CVs were found to be
magnetic intermediate polars (IPs) \citep{barlow06}. Up to now, IPs represent $\sim 5 \%$ of the
\emph{INTEGRAL} sources detected at that energy range and this number is prone to increase in the near
future, as demonstrated by systematic optical follow-ups for some of the $\sim 100$ \emph{INTEGRAL}
unidentified sources \citep[see e.g.][]{masetti08}. From the detection of thousands of discrete low-luminosity
X-ray sources in a deep \emph{Chandra} survey of regions close to the Galactic center, \citet{muno04}
proposed that magnetic CVs of the IP type may represent a significant fraction of the Galactic background.
Further support that magnetic CVs could still be a hidden population of faint X-ray sources and, therefore,
play an important role in the X-ray emission of the Galaxy comes from recent studies of the Galactic ridge
with \emph{INTEGRAL} and \emph{RXTE} \citep{revnivtsev08}.

IPs are believed to harbor weakly magnetized ($B \lesssim 10$ MG) white dwarfs (WDs) because of their
fast asynchronous rotation with respect to the orbital period and of the lack of significant polarized emission
in the optical/near-IR in most systems. This contrasts with the other group of magnetic CVs, the polars,
that instead possess strongly magnetized ($B \sim 10 - 230$ MG) WDs rotating synchronously with the binary
period. The X-ray properties of magnetic CVs are strictly related to the accretion mechanism onto the WD
primary. Material from the late type companion is driven by the magnetic field lines onto the magnetic polar
caps, where a shock develops \citep{aizu73} below which hard X-rays and cyclotron radiation are emitted.
Bremsstrahlung radiation is believed to be the dominant cooling mechanism in IPs \citep{wu94}, while cyclotron
radiation may dominate in polars. The complex interplay between the two mechanisms greatly depends on both
magnetic field strength and local mass accretion rates \citep{woelk_beuermann96,fischer_beuermann01}. Hence,
if IPs indeed host weakly magnetized WDs with respect to polars, this could qualitatively explain why they are
hard X-ray sources. The detection of a soft X-ray optically thick component in an increasing number of systems
\citep{anzolin08} poses further questions in the interpretation of the X-ray emission properties of IPs.
 
In the framework of an ongoing optical identification program, we identified two new members of the IP 
group: 1RXS~J213344.1+510725 = IGR~J21335+5105 (hereafter RXJ2133) \citep{bonnetbidaud06} and
IGR~J00234+6141 = 1RXS~J002258.3+614111 (hereafter IGR0023) \citep{bonnetbidaud07}. Both of them are hard
CVs in the \emph{INTEGRAL} source catalog \citep{bird07}.

The weak hard X-ray source IGR0023 was detected by the \emph{INTEGRAL} satellite during an observation
of the Cassiopeia region of the Galaxy \citep{denhartog06}. The optical counterpart of IGR0023 was
identified by \citet{masetti06}, who proposed a possible magnetic nature of this CV. A tentative $\sim
570$ s optical periodicity was recognized in the $R$ band \citep{bikmaev06}. However, a clear periodic 
modulation of $563.53 \pm 0.62$ s was discovered with optical photometric data by \citet{bonnetbidaud07}
and readily ascribed to the rotational period of the WD, while an orbital period of $4.033 \pm 0.005$
hr was derived from optical spectroscopy. The properties of the \emph{INTEGRAL} spectrum, which is well
fitted by a bremsstrahlung with a temperature of 31 keV, strongly support the magnetic nature of IGR0023.

RXJ2133 was identified as a hard X-ray point source from the \emph{ROSAT} Galactic Plane Survey \citep{motch98}.
A clear persistent optical light pulsation at $570.823 \pm 0.013$ s was then discovered with fast optical
photometry, while optical spectroscopy revealed an additional periodic variability at $7.193 \pm 0.016$ hr
\citep{bonnetbidaud06}. These two periodicities were respectively identified as the WD spin and the orbital
periods, thus suggesting RXJ2133 as a member of the IP class with a relatively long orbital period which
falls into the so-called IP gap between 6.5 and 9.5 hrs \citep{schenker04}. \citet{katajainen07} found also
that RXJ2133 emits optical circularly polarized light up to $\sim 3 \%$ and proposed that the WD magnetic
field could be as high as 25 MG, one of the highest amongst IPs.

The X-ray variability and broad-band spectra of these two systems have been investigated using pointed
\emph{XMM-Newton} \citep{jansen01} observations and publicly available hard X-ray data obtained with the
\emph{INTEGRAL} satellite \citep{winkler03}. In the case of RXJ2133, we also present the temporal and spectral
analysis of a pointed \emph{Suzaku} \citep{mitsuda07} observation.

\section{Observations and data reduction}

The summary of the observations of IGR0023 and RXJ2133 is reported in Table \ref{tab:observ}.

\begin{table*}
\caption{Summary of the X-ray observations of IGR0023 and RXJ2133. We report total exposure times for the
\emph{INTEGRAL} observations and net exposure times for the others.}
\label{tab:observ}
\centering 
\begin{tabular}{l l l l r c}
\hline\hline 
Object	& Instrument	& Date			& UT (start)	& Exposure time (s) & Net count rate ($\ctss$)\\
\hline
IGR0023	& EPIC-pn		& 2007-07-10    & 05:58			& 24\,660		& $0.984 \pm 0.007$ \\
		& EPIC-MOS		&				& 05:36			& 26\,540		& $0.345 \pm 0.004$ \\
		& RGS			&				& 05:35 		& 26\,706		& $0.038 \pm 0.001$ \\
		& OM-V			&				& 05:44			& 1\,960		& $3.26 \pm 0.04$	\\
		&				&				& 06:22			& 1\,959		& $3.12 \pm 0.04$	\\
		&				&				& 07:01			& 1\,960		& $2.79 \pm 0.04$	\\
		&				&				& 08:12			& 1\,962		& $2.97 \pm 0.04$	\\
		&				&				& 08:50			& 1\,959		& $2.85 \pm 0.04$	\\
		& OM-UVM2		&				& 09:28			& 1\,960		& $0.16 \pm 0.01$	\\
		&				&				& 10:06			& 1\,959		& $0.19 \pm 0.01$	\\
		&				&				& 10:45			& 1\,960		& $0.20 \pm 0.01$	\\
		&				&				& 11:23			& 1\,960		& $0.18 \pm 0.01$	\\
		&				&				& 12:01			& 1\,960		& $0.18 \pm 0.01$	\\
& & & & & \\
		& IBIS/ISGRI	& 				&				& $\sim 6\,900\,000$	& $0.10 \pm 0.01$ \\
\hline
RXJ2133 & EPIC-pn		& 2005-05-29	& 13:33			& 13\,585		& $5.42 \pm 0.02$ \\
		& EPIC-MOS		&				& 12:47			& 16\,580		& $1.398 \pm 0.008$ \\
		& RGS			&				& 13:28 		& 13\,890		& $0.147 \pm 0.004$ \\
		& OM-B			&				& 10:11			& 2\,320		& $24.9 \pm 0.1$ \\
		& 				&				& 10:55			& 2\,319		& $23.2 \pm 0.1$ \\
		& 				&				& 11:39			& 2\,319		& $22.4 \pm 0.1$ \\
		& 				&				& 12:23			& 2\,319		& $22.1 \pm 0.1$ \\
		& 				&				& 13:07			& 2\,320		& $23.7 \pm 0.1$ \\
		& OM-UVM2		&				& 13:51			& 2\,319		& $0.69 \pm 0.02$ \\
		& 				&				& 14:35			& 2\,321		& $0.68 \pm 0.02$ \\
		& 				&				& 15:20			& 2\,320		& $0.89 \pm 0.02$ \\
		& 				&				& 16:03			& 2\,320		& $0.85 \pm 0.02$ \\
		& 				&				& 16:48			& 2\,319		& $0.87 \pm 0.02$ \\
 & & & & & \\
		& EPIC-pn		& 2005-07-06	& 18:03			& 9\,871		& $5.12 \pm 0.03$ \\
		& EPIC-MOS		&				& 17:05			& 13\,680		& $1.105 \pm 0.009$ \\
		& RGS			&				& 17:04 		& 13\,910		& $0.120 \pm 0.005$ \\
		& OM-UVM2		&				& 17:31			& 1\,681		& $0.60 \pm 0.02$ \\
		& 				&				& 18:16			& 1\,679		& $0.70 \pm 0.02$ \\
		& 				&				& 19:20			& 1\,680		& $0.64 \pm 0.02$ \\
		& 				&				& 19:53			& 1\,680		& $0.69 \pm 0.02$ \\
		& 				&				& 20:27			& 1\,679		& $0.59 \pm 0.02$ \\
& & & & & \\
		& IBIS/ISGRI	&				&				& $\sim 3\,740\,000$	& $0.55 \pm 0.02$ \\
 & & & & & \\
		& XIS-FI		& 2006-04-29	& 06:50			& 84\,288		& $0.738 \pm 0.002$ \\
		& XIS-BI		&				& 06:50			& 84\,288		& $0.892 \pm 0.004$ \\
		& HXD			&				& 06:50			& 62\,879		& $0.739 \pm 0.004$ \\
\hline
\end{tabular}
\end{table*}

\subsection{The \emph{XMM-Newton} observations}

For all our \emph{XMM-Newton} observations, we reprocessed and analyzed the EPIC-pn \citep{struder01},
MOS \citep{turner01}, RGS \citep{denherder01} and OM \citep{mason01} data using the standard reduction
pipelines included in SAS 8.0 and the latest calibration files. For IGR0023, because of a problem with
the standard source detection task, the OM-$UVM2$ data were reprocessed at MSSL using an unreleased reduction
routine. Heliocentric corrections were applied to the EPIC and OM data of both sources. The SAS tasks \textit{rmfgen}
and \textit{arfgen} were used to generate the photon redistribution matrix and the ancillary region files
for all the EPIC cameras and RGS instruments.

IGR0023 was observed on July 10, 2007 (OBSID: 0501230201) with the EPIC-pn and MOS cameras operated in full frame
imaging mode with the thin and medium filters, respectively. The total exposure times were 25 ks for EPIC-pn and
26.6 ks for both the MOS cameras. The RGS was operated in spectroscopy mode for a total exposure time of 26.9 ks.
The OM was operated in fast imaging mode using sequentially the $V$ (5000--6000 \AA) and $UVM2$ (2000--2800 \AA)
filters, for 9.8 ks each.

A 28\arcsec\ aperture radius was used to extract EPIC light curves and spectra from a circular region
centered on the source and from a background region located on the same CCD chip where the source was imaged.
In order to improve the S/N ratio, we filtered the data by selecting pattern pixel events up to double with zero
quality flag for the EPIC-pn data, and up to quadruple pixel events for the EPIC-MOS data. The average background
level of the EPIC cameras was quite low for almost all the duration of the observation, with the exception of a
moderate flaring activity, which occurred during the last $\sim 2000$ s of the EPIC-pn exposure. This flare did
not significantly affect the data used for the timing analysis, however we conservatively did not consider these
events in the spectral analysis.

Due to the weakness of IGR0023, the RGS data had poor S/N ratios and therefore were not useful for spectral analysis.

Background subtracted OM-$V$ light curves were obtained with a binning time of 20 s, while the OM-$UVM2$ light
curves were provided by MSSL with a binning time of 10 s. The average count rates were $3.00\ \ctss$ in the $V$
band and $0.18\ \ctss$ in the $UVM2$ band, corresponding to instrumental magnitudes $V = 16.8$ and $UVM2 = 17.6$
and average fluxes of $7.4 \times 10^{-16}\ \fluxA$ and $4.0 \times 10^{-16}\ \fluxA$, respectively. As a comparison,
the continuum flux of the optical spectrum obtained by \citet{masetti06} was $\sim 6 \times 10^{-16}\ \fluxA$.

RXJ2133 was observed on May 29, 2005 (OBSID: 0302100101) with all the EPIC cameras operated in full frame
mode and with the medium filter. Due to high background radiation the observation time was shortened and
the net exposure times were 13.6 ks for EPIC-pn, 16.5 ks for EPIC-MOS and 13.9 ks for RGS, the latter operated
in spectroscopy mode. The OM, operated in fast imaging mode, was used sequentially with the $B$ filter,
covering the spectral range 3900--4900 \AA, and the $UVM2$ filter for 11.6 ks each. RXJ2133 was observed again
on July 06, 2005 (OBSID: 0302100301) with the same instrumental configurations and with net exposure times
of 9.9 ks for EPIC-pn, 13.7 ks for EPIC-MOS and 13.9 ks for RGS. The OM was only operated with the $UVM2$
filter for a total exposure time of 8.4 ks. The source was found at about the same count rate at the two
epochs in all instruments.

EPIC light curves and spectra of source and background were extracted from circular regions of 37\arcsec\ radius.
The same filtering adopted for IGR0023 was applied to improve the S/N ratio for both cameras. During the observation
of May 2005 the background was moderately active, but not enough to affect the timing analysis. Filtering of higher
background periods was done only during the extraction of spectra from both EPIC and RGS data. These spectra were
rebinned to have a minimum of 25 and 20 counts per bin, respectively. 

OM $B$ and $UVM2$ light curves were extracted with a binning time of 10 s and 20 s, respectively. The
average count rates were about $25\ \ctss$ in the $B$ band and $0.7\ \ctss$ in the $UVM2$ band,
corresponding to instrumental magnitudes $B = 15.8$ and $UVM2 = 16.2$. These translate into $B$ and $UVM2$
fluxes of $3 \times 10^{-15}\ \fluxA$  and $1.5\times 10^{-15}\ \fluxA$, respectively.

\subsection{The \emph{INTEGRAL} observations}
 
The \emph{INTEGRAL} IBIS/ISGRI \citep{ubertini03,lebrun03} hard X-ray data of both sources
were extracted from all pointings within 12\degr\ from the source positions, spanning from March 2003 to October
2006. The total effective exposure times are $\sim 6.9$ Ms (2875 pointings) and $\sim 3.74$ Ms (1563 pointings)
for IGR0023 and RXJ2133, respectively. To study the weak persistent X-ray emission, the time averaged ISGRI
spectra have been obtained from mosaic images in five energy bands, logarithmic spaced between 20 and 100 keV.
Data were reduced with the standard OSA software version 7.0 and, then, analyzed using the algorithms described
by \citet{goldwurm_etal03}.

\subsection{The \emph{Suzaku} observations of RXJ2133}

RXJ2133 was observed with \textit{Suzaku} between Apr 29, 2006 and May 1, 2006 (sequence number 401038010).
We have analyzed data from the X-ray imaging spectrometer (XIS, \citet{koyama07}) and the non-imaging hard
X-ray detector (HXD, \citet{taka07}). The observation was done with the object at the ``HXD-nominal'' pointing
position, $\sim 5 \arcmin$ off-axis from the center of field-of-view (FOV) of the XIS, to optimize the S/N ratio
of the HXD data. We based our analysis on data processed using the V2.0.6.13 pipeline released as a part of
HEADAS 6.3.1.

For the XIS, we updated the energy scale calibration using the February 1, 2008 release of the calibration
database. We then applied the following screening criteria: attitude control system in nominal mode, pointing
within 1\farcm5 of the mean direction, XIS data rate medium or high, the satellite outside the South Atlantic
Anomaly (SAA) and at least 180 s after the last SAA passage, elevation above Earth limb $>5 \degr$, elevation
about the bright Earth limb $>15 \degr$. An inspection of the XIS image revealed a second source, near the
center of the FOV, with a flux of $6.7 \times 10^{-13}\ \flux$ in the 2--10 keV band. This same source is
seen in the \emph{XMM-Newton} data at similar flux levels~\footnote{We have not investigated the nature of this
second source, but a likely explanation is that it is a background AGN.}. Although being faint, we conservatively
excluded this source from our analysis of the XIS data. We cannot do so for the HXD data analysis, but it is
expected to have a negligible effect.

We used a 3\farcm5 radius extraction region and an annular extraction region of 7\farcm5 outer and 4\arcmin\ inner
radii for the background, both centered on the position of RXJ2133. For spectroscopy, we summed the data and
the responses of the three XIS-FI units because they have nearly identical responses. For photometry, we added
background subtracted light curves from all the XIS units over the energy range 0.3 (FI) / 0.2 (BI) -- 12 keV.
         
For the HXD data, we took the PIN event data from the processing pipeline and applied the dead time correction.
We obtained the ``tuned'' non X-ray background files \citep{fukazawa09}, estimated by the HXD team using LCFITDT
method. For phase-resolved spectroscopy, we used the phase-averaged background and dead-time fraction, since both
tend to vary on a longer time scale.

\section{Data analysis and results}

\subsection{IGR0023}

\subsubsection{The X-ray variability}

\begin{figure}
\centering
\resizebox{\hsize}{!}{\includegraphics{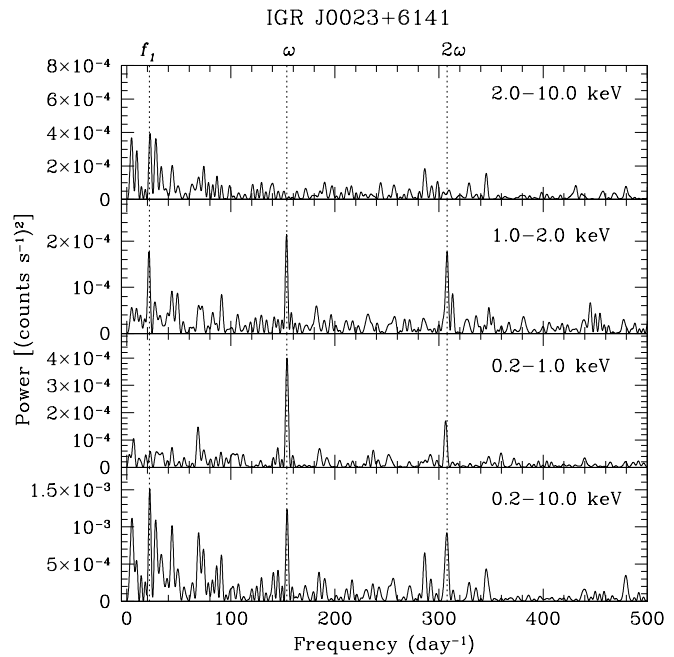}}
\caption{Power spectra of the EPIC-pn data of IGR0023 in selected energy ranges. \textit{From bottom to
top}: 0.2--10.0 keV, 0.2--1.0 keV, 1.0--2.0 keV and 2.0--10.0 keV. Vertical dotted line indicate the
frequencies $\omega$, $2 \omega$ and $f_1$ (see text).}
\label{fig:0023pow}
\end{figure}

The EPIC-pn light curve extracted in the energy range 0.2--10.0 keV and binned in 20 s time intervals
reveals a clear variability with a time scale of the order of $\sim 10$ min. We have not analyzed the EPIC-MOS
light curves because they were too noisy to provide reliable results. The power spectrum of the full-band
EPIC-pn light curve (see Fig. \ref{fig:0023pow}) shows significant peaks at $\omega \sim 154\ \mathrm{d}^{-1}$,
at $2 \omega$ and also at low frequencies. A peak at $f_1 \sim 22\ \mathrm{d}^{-1}$, although its significance
is below $2 \sigma$, is close to a pseudo-periodicity detected in the optical \citep{bonnetbidaud07}.

A sinusoidal fit was performed on the EPIC-pn light curve previously corrected for low frequency trends
by using a third-order polynomial. We used three sinusoids with different frequencies accounting for all the
observed peaks, thus finding $\omega = 153.83 \pm 0.15\ \mbox{d}^{-1}$ and $f_1 = 21.94 \pm 0.15\ \mbox{d}^{-1}$
(errors are at the $1 \sigma$ confidence level). The inferred period $P_\omega = 561.64 \pm 0.56$ s can
be identified with the WD spin period, since the difference with the optical period of \citet{bonnetbidaud07}
is not significant.

A Fourier analysis was also performed on EPIC-pn light curves extracted with a 40 s binning time in selected
energy bands: 0.2--1.0 keV, 1.0--2.0 keV and 2.0--10.0 keV (see Fig. \ref{fig:0023pow}). Peaks at $\omega$
and $2 \omega$ are clearly detected in the low and intermediate energy bands, while they do not appear at
high energies. Instead, the pseudo-periodicity at low frequency is visible only in the intermediate
1--2 keV band.

\begin{figure}
\centering
\resizebox{\hsize}{!}{\includegraphics{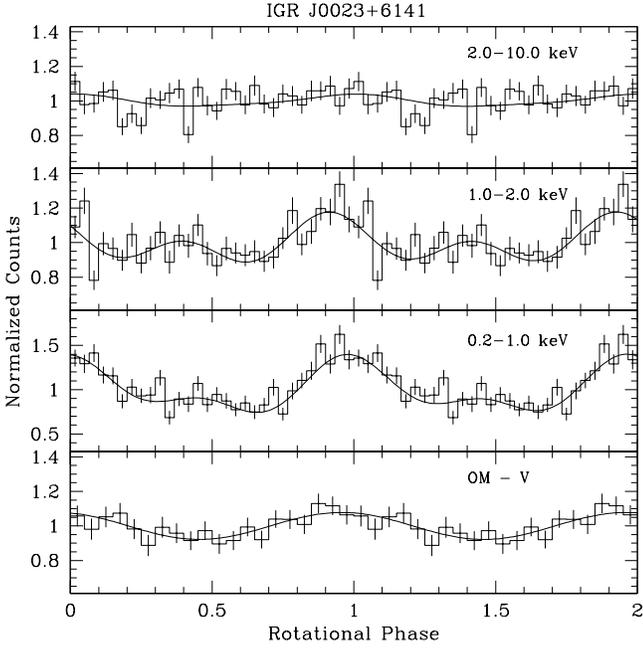}}
\caption{The spin pulse of IGR0023 in selected energy ranges of the EPIC-pn and in the OM-$V$ band.
All the light curves are folded at the 561.6 s X-ray spin period using the time of maximum quoted
in the text. Solid lines represent sinusoidal fits to the corresponding data.}
\label{fig:0023flc}
\end{figure}

We then folded the EPIC-pn light curves with the 561.64 s spin period (Fig. \ref{fig:0023flc}) using the
time of maximum obtained from the sinusoidal fit of the 0.2--10.0 keV light curve: $HJD = 2454291.8667(2)$.
These light curves show a quasi-sinusoidal modulation only below 2.0 keV, with a secondary maximum
at $\phi \sim 0.5$. The pulse amplitude is $\sim 50 \%$ in the 0.2--1.0 keV range and $\sim 16 \%$ in the
1.0--2.0 keV range. The count ratio between the 1.0--2.0 keV and the 0.2--1.0 keV bands indicates an hardening
of the emission at the secondary maximum.

\subsubsection{The visible and UV light curves}

The power spectrum of the OM-$V$ light curve shows a strong peak at the X-ray spin period, while that of
the $UVM2$ light curve does not reveal any significant peak. The average count rate is not constant in
the 5 observations with the $V$ filter spanning $\sim 3/4$ of the orbital cycle, probably suggesting a
dependence on the orbital period. The spin-folded OM-$V$ light curve (lower panel of Fig.\ref{fig:0023flc})
is pretty sinusoidal, with an amplitude of $8 \pm 1 \%$, and shows a single peak at
a phase consistent with that of the main maximum of the X-ray pulse. We notice that \citet{bonnetbidaud07}
also found a single peaked pulsation in their observations carried out with the Gunn-$g$ filter.

\subsubsection{Spectral properties of IGR0023}

\begin{figure}
\centering
\resizebox{\hsize}{!}{\includegraphics{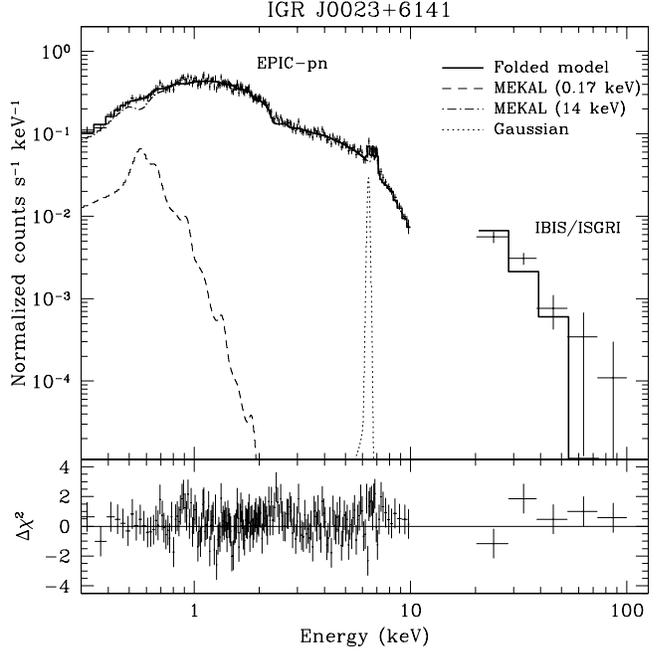}}
\caption{The EPIC-pn and \emph{INTEGRAL} spectra of IGR0023 are shown with model B discussed in the text and
parameters reported in Table \ref{tab:spectra0023} (we do not report the combined MOS spectrum for clarity).
Each spectral component that contributes to the best-fit model is shown separately. The bottom panel shows the
residuals expressed in terms of $\sigma$.}
\label{fig:0023spec}
\end{figure}

The EPIC-pn and combined MOS spectra (in the range 0.3--10.0 keV) and the IBIS/ISGRI spectrum (in the range
20--100 keV) were simultaneously analyzed with the XSPEC 12 package. An absorbed isothermal optically
thin MEKAL component plus a zero width Gaussian at 6.4 keV fit relatively well the spectrum ($\chi_{\nu}^2
= 1.04$), but the temperature is unconstrained ($> 78$ keV). The fit improves using a multi-temperature CEMEKL
emission component and a dense ($N_H \sim 10^{23}\ \mathrm{cm}^{-2}$) absorber covering $\sim 40 \%$ of the
source (model A in Table \ref{tab:spectra0023}), although the $\alpha$ parameter assumes an unreasonably
high value. Metal abundances are consistent, within errors, with the solar values.
The total absorber is likely of interstellar origin, as it is comparable
to that of the total Galactic absorption in the direction of the source ($N_{\mathrm{H, gal}} = 7.4 \times
10^{21}\ \mathrm{cm}^{−2}$, \citet{dickeylockman90}). The dense partial absorber, instead, is likely located
close to the source, as suggested by the energy dependence of the spin light curve. We also obtained similar
quality fits by substituting the CEMEKL component with 2 (Fig. \ref{fig:0023spec}) or 3 MEKALs (model B and C
in Table \ref{tab:spectra0023}, respectively). In both cases,
we find a low ($k T \sim 0.17$ keV) and an intermediate temperature ($k T \sim 10$ keV) components, and, in
model C, a lower limit to the temperature of the third MEKAL $k T > 27$ keV. The observed X-ray flux in the
0.2--10.0 keV range is $6.8 \times 10^{-12}\ \flux$. We notice that only model B
and model C account for the \ion{O}{VII} (21.9 \AA) and \ion{O}{VIII} (19.1 \AA) lines barely detectable in
the low quality RGS spectra.

The \emph{INTEGRAL} spectrum above 60 keV seems to be underpredicted by the three best-fit models but,
given the poor S/N ratio, the excess of counts is not significant. The addition of a reflection component,
suggested by the presence of the Fe line at 6.4 keV, does not improve the fits. We also fitted
the broad-band spectrum of IGR0023 with the recent post-shock region (PSR) model of \citet{suleimanov08}
that computes the emergent spectrum taking also into account the Compton scattering ($\chi^2$ / d.o.f. = 136/110).
We obtained a shock temperature $k T_{\mathrm{shock}} = 51 \pm 11$ keV and absorption parameters consistent with
those found with models A, B and C. The contribution of the Compton scattering is found to be $\sim 10 \%$
and, within errors, does not seem to affect the temperature determination.

A phase-resolved analysis of the EPIC-pn spectrum of IGR0023 was performed selecting  two phase intervals
centered on the two maxima of the spin pulse ($\phi = 0.85 - 1.15$ and $\phi = 0.35 - 0.65$, respectively).
We used model B, since model A and C would give badly constrained parameters. The hydrogen column density of
the total absorber, the temperatures of the two MEKALs and the metal abundance were kept fixed to the values
obtained for the average spectrum. As shown in Table \ref{tab:phspectra0023}, the normalization of the
low-temperature optically thin component is significantly lower at the secondary maximum, where we also find
a marginal evidence of an increase of the covering fraction of the local absorber.

\begin{table}
\caption{Spectral parameters obtained from fitting the EPIC-pn spectra of IGR0023 extracted in the
phase intervals quoted in the text with model B shown in Table \ref{tab:spectra0023}.} 
\label{tab:phspectra0023}
\centering
\begin{tabular}{lll}
\hline\hline
Parameters 					  	& Maximum 1					& Maximum 2 \\
\hline
$N_{\mathrm{H}}$ ($10^{23}\ \mathrm{cm}^{-2}$) & $1.5_{-0.6}^{+1.0}$ & $1.2_{-0.3}^{+0.5}$ \\
Cov. Frac.						& $0.36_{-0.07}^{+0.06}$	& $0.51 \pm 0.04$ \\
$C_1$ ($10^{-4}$)				& $1.2_{-0.6}^{+0.3}$ 		& $0.5 \pm 0.5$	\\
$C_2$ ($10^{-2}$)				& $1.5_{-0.1}^{+0.2}$ 		& $1.7 \pm 0.1$	\\
\hline
$F_{0.2-10.0\ \mathrm{keV}}$	& 2.06						& 2.00 \\
($10^{-12}\ \flux$) 			&	& \\
\hline
$\chi_{\nu}^2$ ($\chi^2$ / d.o.f.) & 1.01 (259/257)			& 1.10 (258/235) \\
\hline
\end{tabular}
\end{table}

\subsection{RXJ2133}

\subsubsection{The X-ray periodicities in RXJ2133}

\begin{figure}
\centering
\resizebox{\hsize}{!}{\includegraphics{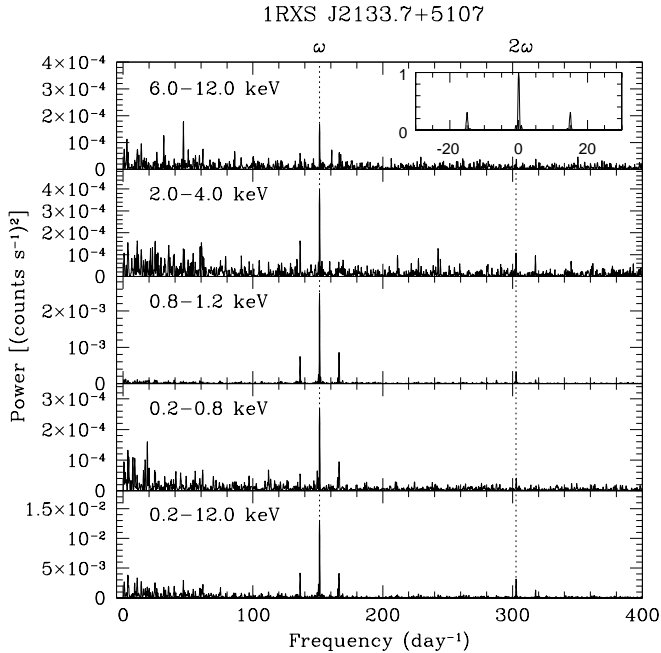}}
\caption{Power spectra of the \emph{Suzaku} XIS light curves of RXJ2133 in selected energy ranges.
\textit{From bottom to top}: 0.2--12.0 keV, 0.2--0.8 keV, 0.8--1.2 keV, 2.0--4.0 keV and 6.0--12.0 keV.
Vertical dotted lines indicate the frequencies $\omega$ and $2 \omega$ (see text). The spectral window
is shown in the inset.}
\label{fig:2133pow}
\end{figure}

The EPIC light curves, extracted in the 0.2--12.0 keV energy range and binned in 5 s time intervals,
and the \emph{Suzaku} XIS light curve, extracted with a binning time of 8 s, show a variability of
the order of 10 min. In the EPIC light curve of May 2005 a quasi-sinusoidal trend with an apparent period
of $\sim 2$ hr is found, while in July 2005 a $\sim 1$ hr pseudo-periodicity is detected. However, in the later
XIS observation we do not find any evidence of such a variability.

We then performed a Fourier analysis on the light curves at the three epochs. Peaks at the optically
identified spin frequency, $\omega$, and at $2\omega$ are clearly detected. In Fig. \ref{fig:2133pow} we report
the full-band XIS power spectra, as well as that obtained in different
energy bands. Sinusoidal fits to the full-band light curves give $\omega = 151.30 \pm 0.25\ \mbox{d}^{-1}$,
$\omega = 151.55 \pm 0.78\ \mbox{d}^{-1}$ and $\omega = 151.350 \pm 0.009\ \mbox{d}^{-1}$ for the May and
July 2005 \emph{XMM-Newton} and April 2006 \emph{Suzaku} data sets, respectively. In the analysis of the
EPIC-pn light curves, a third sinusoid has to be included, accounting for the low frequency variations, thus
giving $f_{\mathrm{May 2005}} = 43.78 \pm 0.29\ \mbox{d}^{-1}$ and $f_{\mathrm{July 2005}} = 27.87 \pm
0.41\ \mbox{d}^{-1}$. We identify the precise \emph{Suzaku} period $P_\omega = 570.862 \pm 0.034\ \mathrm{s}$ with
the true spin period of the accreting WD. This agrees, within errors, with the optical $570.823 \pm 0.013$ s period. 

\begin{figure}
\centering
\resizebox{\hsize}{!}{\includegraphics{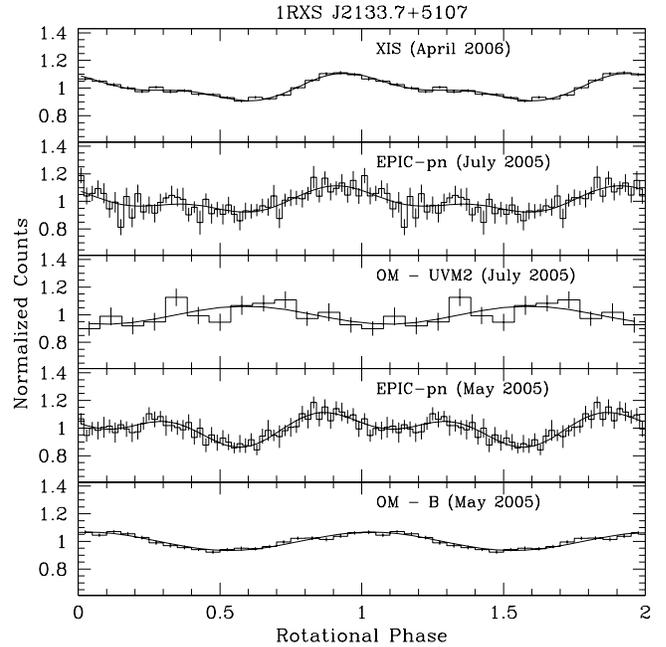}}
\caption{Spin-folded light curves of RXJ2133. \textit{From top to bottom}: full-band XIS of April 2006,
EPIC-pn and OM-$UVM2$ of July 2005, EPIC-pn and OM-$B$ of May 2005. Solid lines represent sinusoidal fits
to the corresponding data.}
\label{fig:2133flc1}
\end{figure}

We folded the EPIC-pn and XIS light curves at the 570.86 s X-ray spin period using the time of maxima of the
pulsation at $\omega$: $HJD_{\mathrm{May 2005}} = 2453520.1444(2)$, $HJD_{\mathrm{July 2005}} = 2453558.3134(4)$
and $HJD_{\mathrm{April 2006}} = 2453856.09618(5)$. The three folded light curves (Fig. \ref{fig:2133flc1})
present two maxima at phases 0.9 and 0.35, with a dip at phase $\sim 0.1$ that is more evident
in the May 2005 data. The full amplitude of the primary maximum at $\phi \sim 0.9$ is almost similar at
the three epochs ($\sim 36 \%$), while that of the secondary maximum decreases from $\sim 13 \%$ in May 2005
to $\sim 3\%$ in 2006.

\begin{figure}
\centering
\resizebox{\hsize}{!}{\includegraphics{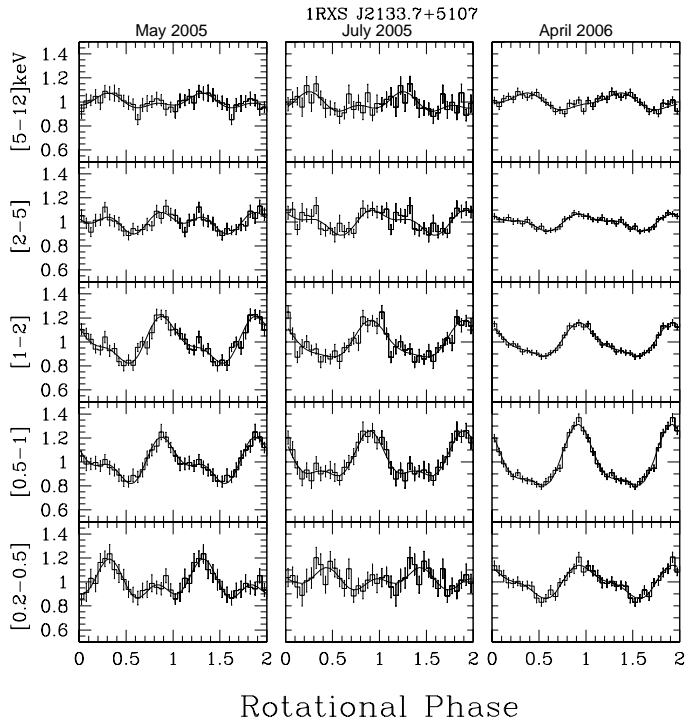}}
\caption{EPIC-pn (May 2005 and July 2005) and XIS  spin-folded light curves of RXJ2133 
extracted in selected
energy ranges. All the light curves are folded  using the time of maximum quoted in 
the text.}
\label{fig:2133flc2}
\end{figure}

The energy-resolved EPIC-pn light curves (Fig. \ref{fig:2133flc2}) generally show the
secondary maximum dominating the emission in the soft 0.2--0.5 keV band and the primary maximum dominating in
the range 0.5--2.0 keV. At higher energies the two maxima have similar amplitudes of $\sim 10 \%$. The XIS
light curves, though being broadly similar to the EPIC-pn light curves, present an evident secondary maximum
only in the hard X-rays.

\subsubsection{The UV and optical light variability of RXJ2133} \label{sec:om2133}

The power spectrum of the OM-$B$ light curve shows a strong peak at the spin frequency (consistent
within errors with the X-ray and the previous optical determinations).
We also detect power at low frequencies, but we cannot establish whether this indicates a true periodic
variability. The OM-$B$ light curve of May 2005 folded at the spin period (see Fig. \ref{fig:2133flc1})
is single peaked, with a broad maximum at phase $\sim 0$ and a full amplitude of $14 \%$.
The UV light curve, instead, is almost unmodulated in May 2005, while in July 2005 it is sinusoidal
with a maximum shifted by half a period with respect to the OM-$B$ light curve of May 2005.
A similar antiphased behavior is also seen in some soft X-ray IPs, like UU Col \citep{demartino06a}.
However, we point out that each OM light curve covers only $\sim 41\%$ of the orbital period and hence
could be affected by orbital dependent changes of the spin modulation.

\subsubsection{The X-ray spectrum of RXJ2133}

\begin{figure}
\centering
\resizebox{\hsize}{!}{\includegraphics{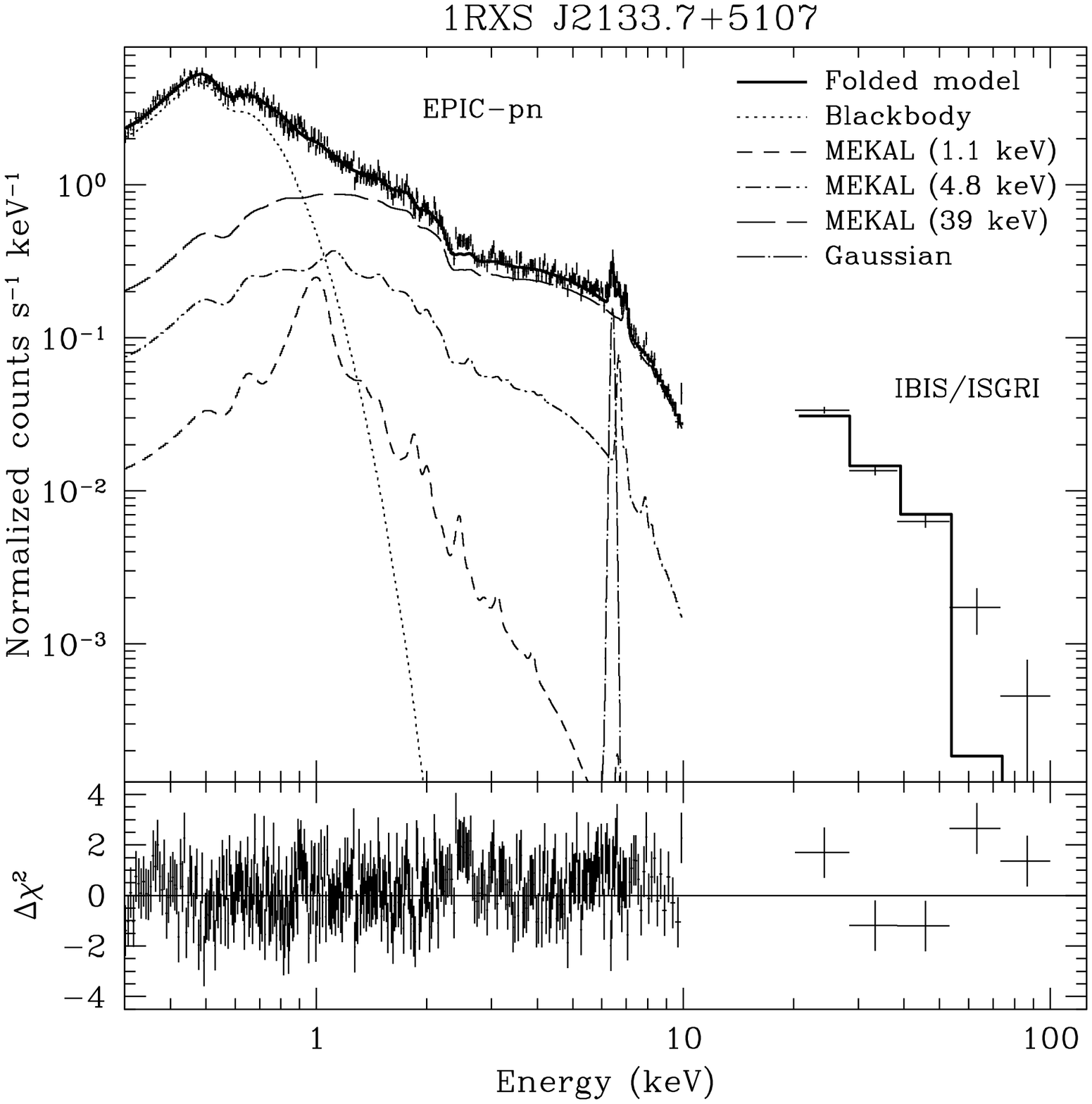}}
\resizebox{7.5cm}{!}{\includegraphics{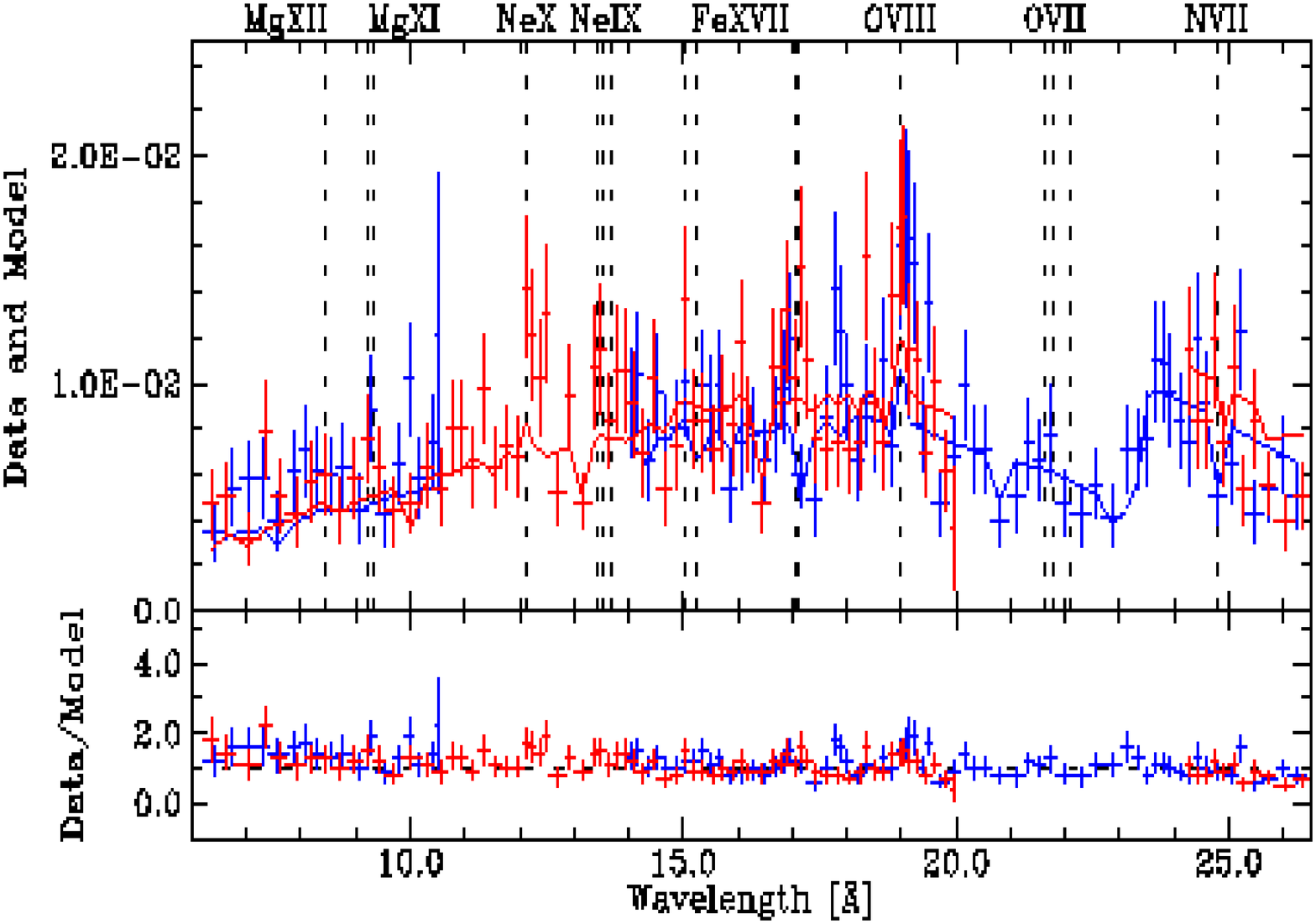}}
\caption{\textit{Upper figure}: the EPIC-pn (May 2005) and \emph{INTEGRAL} spectra of RXJ2133 are shown with
model C discussed in the text and parameters reported in Table \ref{tab:spectra2133} (the combined EPIC-MOS
spectrum is not plotted for clarity). The different spectral components contributing to the best-fit model
are shown separately. Bottom panel shows the residuals expressed in terms of $\sigma$. \textit{Lower figure}:
the RGS spectra of RXJ2133.}
\label{fig:2133spec}
\end{figure}

\begin{figure}
\centering
\resizebox{\hsize}{!}{\includegraphics{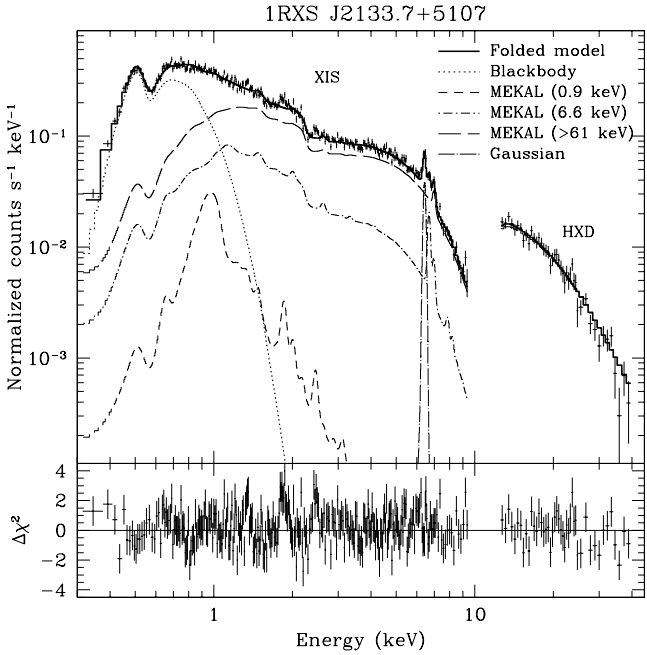}}
\caption{The \emph{Suzaku} spectra of RXJ2133 are shown with model C discussed in the text and
parameters reported in Table \ref{tab:spectra2133}. The bottom panel shows the residuals.}
\label{fig:2133specsuz}
\end{figure}

The X-ray spectrum of RXJ2133 was studied in the broad-band energy range 0.3--100 keV using the EPIC-pn and
combined MOS data at each epoch, together with the IBIS/ISGRI spectrum. We also analyzed the combined XIS-FI,
XIS-BI and HXD spectra in the restricted range 0.3--40 keV, where the S/N ratio is still good. The average X-ray
spectrum is well fitted by a multi-temperature optically thin component plus a rather hot ($k T_{\mathrm{BB}}
\sim 100$ eV) blackbody, that is required to best fit the data at energies below 1 keV.
A total absorber with a low hydrogen column density, likely of interstellar origin (the total
Galactic absorption along the direction of the source is $N_{\mathrm{H, gal}} = 7.35 \times 10^{21}\ \mbox{cm}^{-2}$,
\citet{dickeylockman90}), and a dense ($N_H \sim 10^{23}\ \mbox{cm}^{-2}$) absorber partially covering the source
are also required. In all the spectral fits we fixed the $N_H$ of the total absorber to the value found for
the May 2005 data, since it is not expected to vary with time, and used an unresolved Gaussian line to account
for the fluorescence Fe line at 6.4 keV. The best-fit parameters are reported in Table \ref{tab:spectra2133},
where the optically thin components in models A, B and C are the same used in the corresponding spectral models
of IGR0023. We have marked with \textit{1} and \textit{2} the May and July 2005 \emph{XMM-Newton} spectra,
respectively, while \textit{3} indicates the \emph{Suzaku} spectra~\footnote{We caution that the quoted errors
reflect only the statistical errors. However, our experimentation of adjusting the HXD non X-ray background by
$\pm 1\%$ from the best estimate values did not cause significant changes in the fit parameters, showing that the
systematic errors are not dominant in this case.}. Also for this object, the \emph{INTEGRAL} spectrum is
underpredicted above $\sim 60$ keV.

In general, we find no significant differences between the spectral parameters of May 2005 and July 2005, with
the exception of the normalization of the blackbody component that is slightly higher in the earliest epoch.
The absorbed flux in the range 0.2--10.0 keV is $2.4 \times 10^{-11}\ \flux$. In the \emph{Suzaku} spectra,
the temperature of the hottest MEKAL (or the maximum temperature of the CEMEKL model) attains higher values than
in 2005. In addition, we find a decrease of the covering fraction of the partial absorber $(50 \%)$ and a lower
value of the normalization of the blackbody (with the exception of model C). The absorbed flux in the 0.2--10.0
keV range is $2.5 \times 10^{-11}\ \flux$, in agreement with that found in the two observations of 2005.
Although we find different values of the bolometric fluxes of both the blackbody and the optically thin
components between the two epochs, they are consistent within errors:
$F_{\mathrm{BB}} = (4.1_{-0.4}^{+0.5} - 4.8_{-0.8}^{+0.6}) \times10^{-11}\ \flux$ and
$F_{\mathrm{hard}} = (7.3_{-1.4}^{+1.1} - 7.6_{-0.9}^{+0.8}) \times10^{-11}\ \flux$
in 2005, $F_{\mathrm{BB}} = 4.0_{-0.6}^{+1.0} \times 10^{-11}\ \flux$ and
$F_{\mathrm{hard}} > 8.3 \times 10^{-11}\ \flux$ in 2006.
Therefore, the ratio between the soft and hard X-ray fluxes is of the order of 0.5.

Because of the similar temperatures of the optically thin components of model C found at the three epochs,
we adopt this model to describe the broad-band spectrum of RXJ2133 (see Fig. \ref{fig:2133spec} and Fig.
\ref{fig:2133specsuz}). Although there is clear evidence of a temperature gradient in the post-shock region,
the broad-band spectra do not provide constraints on lower temperatures.
Instead, the presence of a low temperature plasma can be inferred from the May 2005 RGS spectra, where the strong
\ion{O}{VIII} ($0.651 \pm 0.002$ keV , $E.W. = 22$ eV) and \ion{Ne}{X} ($1.02 \pm 0.01$ keV, $E.W. = 14$ eV)
lines are clearly visible (see Fig. \ref{fig:2133spec}). The \ion{O}{VII} line at 0.58 keV ($E.W. = 8$ eV)
and the \ion{Fe}{XVII} line at 0.73 keV ($E.W. = 11$ eV) are likely present, while there is only a weak
evidence of the \ion{Ne}{IX} line. We then estimate a ratio $\sim 2.5$ between the H- and He-like oxygen
lines, that would imply a temperature of  $\sim 0.3-0.4$ keV. This is also supported by the Ne line ratio
of $\sim 1.5$, again indicating $k T \sim 0.4$ keV.

The Fe line at 6.4 keV has a large equivalent width ($E.W. = 150 - 170$ eV), suggesting a Compton reflection
component. However, we do not find improvements of the fits with the inclusion of a reflection
component. We then applied the PSR model to the combined \emph{XMM-Newton} (May 2005) and \emph{INTEGRAL}
spectra ($\chi^{2}$ / d.o.f. = 347/279), as well as to the \emph{Suzaku} spectra ($\chi^{2}$ / d.o.f. = 395/359).
We find a shock temperature of $50 \pm 2$ keV and $53 \pm 3$ keV, respectively,
while the parameters of the absorbers and of the blackbody component are consistent, within errors,
with those found using models A, B and C.

We also performed a phase-resolved analysis of the EPIC-pn (May 2005) and XIS spectra. The former was
extracted in phase intervals approximately centered on the two maxima ($\phi = 0.72 - 0.94$
and $\phi = 0.16-0.28$) and at the minimum ($\phi = 0.44-0.61$). Instead, the higher quality XIS spectrum was
extracted in 4 consecutive phase intervals, namely $\phi = 0.05-0.3$ (P1), $0.3-0.55$ (P2), $0.55-0.8$
(P3) and $0.8-1.05$ (P4). We fitted these spectra adopting model C, fixing $N_H$ of the total absorber,
the temperatures of all the emission components and $A_Z$ to the values found for the corresponding average
spectrum. The results reported in Table \ref{tab:phspectra2133} show significant differences in some of the
parameters. The normalization of the soft MEKAL increases at the primary maximum in the two epochs, whilst
only upper limits are found at other phases. Also, a decrease of the hydrogen column density of the partial
absorber and a slight increase of the blackbody normalization at the secondary maximum are found
only in May 2005.

\begin{table*}
\caption{Spectral parameters obtained from fitting model C shown in Table \ref{tab:spectra2133} to the
EPIC-pn (May 2005) and XIS spectra of RXJ2133 extracted in the phase intervals quoted in the text.}
\label{tab:phspectra2133}
\centering
\begin{tabular}{l|lll|llll}
\hline\hline
 & \multicolumn{3}{c|}{EPIC-pn (May 2005)} & \multicolumn{4}{c}{XIS (April 2006)} \\
\hline
Parameters & Maximum 1 & Maximum 2 & Minimum & P1 & P2 & P3 & P4\\
\hline
$N_{\mathrm{H}}^{\mathrm{PCFABS}}$ ($10^{23}\ \mathrm{cm}^{-2}$) & $1.4_{-0.4}^{+0.6}$ & $0.6 \pm 0.2$
& $1.2_{-0.4}^{+0.6}$ & $1.1 \pm 0.2$ & $1.0 \pm 0.1$ & $1.1 \pm 0.2$ & $1.1 \pm 0.2$ \\
$C_F$ & $0.45_{-0.06}^{+0.05}$ & $0.56_{-0.07}^{+0.06}$ & $0.55_{-0.07}^{+0.05}$ & $0.44 \pm 0.02$ &
$0.53 \pm 0.01$ & $0.46 \pm 0.02$ & $0.38 \pm 0.02$ \\
$C_{\mathrm{BB}}$ ($10^{-4}$) &  $4.7 \pm 0.5$ & $6.4_{-0.9}^{+1.0}$ & $4.6_{-0.6}^{+0.7}$ &
$4.7 \pm 0.2$ & $4.9 \pm 0.2$ & $4.2 \pm 0.2$ & $4.4 \pm 0.2$ \\
$C_1$ ($10^{-4}$) & $14 \pm 3$ & $< 2.5$ & $< 4.2$ & $< 0.8$ & $< 0.2$ & $< 1.5$ & $6.4 \pm 0.8$ \\
$C_2$ ($10^{-3}$) & $< 2.3$ & $5.2_{-2.6}^{+2.7}$ & $2.6_{-2.2}^{+2.5}$ & $3.6_{-0.7}^{+0.8}$ &
$3.9_{-0.6}^{+0.5}$ & $2.7 \pm 0.7$ & $3.7_{-0.7}^{+0.6}$ \\
$C_3$ ($10^{-2}$) & $1.6 \pm 0.2$ & $1.2 \pm 0.2$ & $1.4 \pm 0.2$ & $1.48_{-0.08}^{+0.09}$ &
$1.55_{-0.06}^{+0.07}$ & $1.47_{-0.08}^{+0.09}$ & $1.32 \pm 0.08$ \\
\hline
$F_{0.2-10.0\ \mathrm{keV}}$ & 2.51 & 2.38 & 2.12 & 2.52 & 2.48 & 2.35 & 2.47 \\
($10^{-11}\ \flux$) & & & & & & & \\
\hline
$\chi_{\nu}^2$ & 1.04 & 0.96 & 0.97 & 0.82 & 0.87 & 0.80 & 0.87 \\
($\chi^2$ / d.o.f.) & (434/418) & (266/276) & (280/290) & (1651/2019) & (1721/1982) &
(1548/1939) & (1747/2013) \\
\hline
\end{tabular}
\end{table*}

\section{Discussion}

\subsection{IGR0023}

The \emph{XMM-Newton} data confirm the previously identified optical 561 s period as the spin period of
the accreting WD. A pseudo-periodicity of the order of 1 hr, though with a significance level below $3 \sigma$,
seems to be present above 1 keV and was previously found also in optical data \citep{bonnetbidaud07}.
This variability, if real, is not easy to explain. Assuming a circular Keplerian orbit, a periodicity of
1.09 hr implies a radius of $\sim 3 \times 10^{10}$ cm. For comparison, adopting a mass ratio $q = 0.5$ and a
WD mass of $\sim 0.9 M_{\sun}$ (see below), the outer edge of the accretion disc would be  $4 \times 10^{10}$ cm.
This may suggest an origin within the disc, but unlikely due to absorption effects as instead found in many IPs
\citep{parker_etal05}.

The spin pulse is detected only below 2 keV, where the power spectrum shows substantial 
signal also at $2 \omega$. The double-humped spin pulse and the increase of amplitudes with 
decreasing energy could be the signatures of two accreting poles, as well as of variable absorption.
From the phase resolved spectra, we find a decrease by a factor $\sim 2$ in the normalization constant of
the MEKAL at 0.17 keV, as well as an increase of a dense partial absorber at the secondary maximum. 
This prevents us to isolate the true contribution from a secondary pole. In addition, the 
optical pulse is single-humped, thus not revealing a secondary pole. Hence, we are left with 
the uncertainty on whether the X-ray pulse shape is due to the presence of a secondary and less 
active pole or to phase-dependent absorption effects in the accretion curtain \citep{rosen88}.

The X-ray emission in IGR0023, extending up to $\sim 90$ keV, requires a multi-temperature optically thin
plasma with a temperature distribution more complex than a simple power law. The spectral fits indicate a likely
maximum temperature of $\sim 50$ keV. High plasma temperatures were also inferred from broad-band
analysis of the combined \emph{XMM-Newton -- INTEGRAL} spectra of 1RXS~J173021.5-055933
\citep{demartino08} and from hard X-ray observations of several other IPs \citep{landi08,suleimanov08,brunschweiger09}.
Although caution has to be taken to interpret these temperatures in terms of the WD mass, 
the extension of spectra above 30 keV allow us to directly observe the exponential cut-off of the
underlying continuum and hence the inferred temperatures are likely to be more reliable than those 
based only on softer energy coverages (e.g. \emph{XMM-Newton}). For IGR0023 we estimate a WD mass of 
$0.91^{+0.14}_{-0.16} M_{\sun}$, consistent with that found by \citet{brunschweiger09} from the analysis of
\emph{Swift/BAT} data. Although relatively massive WD primaries were inferred in several magnetic CVs
\citep[see e.g.][]{ramsay00,brunschweiger09}, the mass distribution of magnetic CV primaries is not very much
different from that of non-magnetic systems \citet{ramsay00}. Hence, for a CV to be a hard X-ray source, there
should be an additional parameter that plays an important role.

The hard X-ray bolometric flux $F_{\mathrm{hard}} = 1.8 \times 10^{-11}\ \flux$ is an 
approximate estimate of the total accretion luminosity, as the pulsed components at other wavelengths
need to be included as well as a secure distance estimate is missing. Here we can obtain 
only a lower limit using the 8 \%  pulsed $V$ band contribution, adopting $E(B-V) = 0.35$ as derived by the
$N_H$ of the total absorber \citep{ryter75}. Also, assuming a lower limit
of 500 pc for the distance of IGR0023, as estimated by \citet{bonnetbidaud07} using the 2MASS $K$ band magnitude
when attributed solely to the secondary star \citep{knigge06}, we obtain $L_{\mathrm{accr}} > L_{\mathrm{hard}}
+ L_V = 5.5 \times 10^{32}\ \lum$.
This in turn gives us a lower limit for the accretion rate $\dot{M} > 2.5 \times 10^{15}\ \mdot$. Unless the
distance is much larger and the hard X-rays do not trace the bulk of accretion, this value is much lower than
$2.2 \times 10^{17}\ \mdotn$ predicted for its 4.033 hr orbital period \citep{warner}.

The short spin periods of IGR0023 and RXJ2133 are similar to those of YY Dra \citep{patterson92}, V405 Aur
\citep{skillman96} and 1RXS~J070407.9+262501 \citep{gan05}. Fast rotators were proposed to possess weakly
magnetized WDs \citep{norton99}, although V405 Aur and RXJ2133 are among the few IPs showing polarized
optical emissions. Assuming a  pure disc accretion, as indicated by the absence of orbital sidebands in
the power spectrum, a lower limit for the magnetic moment $\mu \gtrsim 9 \times 10^{31}\ \magmom$ is 
obtained using the inferred values of $M_{\mathrm{WD}}$ and $\dot{M}$. This 
could suggest that IGR0023 is not strongly magnetized and not spinning at equilibrium.

\subsection{RXJ2133}

Our X-ray analysis confirms the 571 s period detected in optical photometry and polarimetry
\citep{bonnetbidaud06,katajainen07} as the rotational period of the WD primary. RXJ2133 is, therefore, one of
the most asynchronous systems among IPs, with a ratio between the spin and the orbital periods of 0.022.

The double peaked X-ray pulse shows a complex energy dependence and might be interpreted in terms of
absorption and two emitting poles. The energy-resolved light curves suggest that the primary pole dominates the
emission in the energy range 0.5--5.0 keV, while the two poles contribute similarly at higher energies. This is
also found in the phase-resolved spectral analysis where the main changes are due to the emitting volume of an
intermediate temperature region. The soft X-ray blackbody component is seen
at all spin phases, suggesting either that the irradiated main pole is never occulted or that reprocessing
occurs at both poles. However, in 2005 the secondary maximum is dominant in the soft band, and at this phase
we find a decrease in the absorption and a slight increase in the blackbody normalization.
Hence, when the main pole points away from the observer, we are viewing the contribution of the irradiated
area of the secondary pole. In order to have a lower absorption at this phase, a modification of the standard
accretion curtain model \citep{norton99}, where the optical depth is 
largest when viewing the curtain perpendicularly, could be envisaged.

UV and optical data also support a scenario with two accreting poles whose 
contributions are however variable with time. The \emph{XMM-Newton} UV observation of July 2005 
shows a weak modulation antiphased with the X-ray pulse, while in May 2005 the UV light is not 
modulated but the $B$ band pulse is single-peaked. Previous white light photometry acquired in 2003 
\citep{bonnetbidaud06} shows the presence of two poles, being the first harmonic of both the spin 
frequency and the beat clearly detected. This is not seen in the 2006 optical photometry and polarimetry 
\citep{katajainen07} as well as in the \emph{Suzaku} X-ray data in the same year.
Note that \citet{bonnetbidaud06} suggested a relatively low binary
inclination $(i \leq 45 \degr)$ and the behavior of the circular polarization along the spin 
cycle indicates a magnetic colatitude $\beta \sim 90\degr - i$ \citep{katajainen07}. This implies $i \geq 
45\degr$ and hence that the secondary pole, whenever active, can be observed.

The broad-band X-ray spectrum of RXJ2133 is well fitted by a multi-temperature optically thin plasma
with a likely maximum temperature $k T_{\max} \sim 50$ keV. As for IGR0023, a 
power-law multi-temperature flow appears a too simple and inadequate 
description of the post-shock emitting region, implying that the emergent spectrum is highly 
sensitive to local pressure and temperature across the flow. Furthermore, a blackbody component
at 100 eV is also required to account for the soft X-ray emission. To date this component has
been detected in $\sim 42 \%$ of the IPs, thus appearing a common characteristics of magnetic CVs and not solely 
of polars \citep{anzolin08}. If produced by the heating of a small area 
surrounding the WD polar cap, the range of values of the normalization constant obtained by 
applying model C to the spectra of RXJ2133 (see Table \ref{tab:spectra2133}) implies an emission area of
$1.6 - 1.9 \times 10^{13}\ d_{\mathrm{600 pc}}^2\ \mbox{cm}^2$. Here we use a minimum distance to the source
as derived by \citet{bonnetbidaud06} using 2MASS $K$ band magnitude. Assuming  $k T_{\max} \sim k T_{\mathrm{shock}}$,
we infer a WD mass of $0.93 \pm 0.04 M_{\sun}$, in agreement with the values found by \citet{bonnetbidaud06}
and \citet{brunschweiger09}, and a radius  of $5.6 \times 10^8$ cm \citep{nauenberg72}. The 
fractional area of the blackbody emitting region is $f \sim 2 - 3 \times 10^{-6}\ d_{\mathrm{600 pc}}^2$.
This could represent a small core onto the WD pole, as found in other soft X-ray IPs with similar temperatures
\citep{haberl02,demartino08}. We also point
out that a $\sim 100$ eV blackbody would imply that the emission from the accretion region locally exceeds the
Eddington limit. At that temperature, the radiation pressure is 4 times stronger than the gravity at the surface
of a $0.93 M_{\sun}$ WD. Clearly we still need a better understanding of emission mechanisms to solve this puzzle.

The bolometric flux of the blackbody represents $\sim 40 \%$ of the total bolometric X-ray emission.
RXJ2133 shows circularly polarized emission up to $\sim 3 \%$ and hence the
flux due to cyclotron radiation cannot be neglected. Since the blackbody flux does not exceed that 
of the hard X-ray emission, the reprocessed radiation seen in the soft X-rays does not balance the primary
radiation components. We have found that the UV emission is modulated by
$6 \%$ at the spin period in July 2005, but antiphased with respect to the hard X-ray component, thus suggesting
that the reprocessed radiation at the WD surface is also emitted at UV wavelengths. The pulsed UV flux 
observed in July 2005, dereddened for $E(B-V) = 0.25$, is  $F_{\mathrm{UVM2}} \sim 2.1 \times 10^{-15}\ \fluxA$.
This value is much larger than that expected from the extrapolation of the soft X-ray flux towards 
UV wavelengths. Also, the UV flux integrated over the $UVM2$ band is only 4 \% of the soft 
X-ray bolometric flux and hence only provide a lower limit to the UV luminosity.

To evaluate the mass accretion rate we then consider the hard and soft X-ray components, as well as the modulated
fraction of the flux in the $B$ band of the May 2005 observation. A lower limit for the accretion luminosity is
therefore: $L_{\mathrm{accr}} \gtrsim 5 \times 10^{33}\ \lumrxj$. We obtain  $\dot{M} \gtrsim 2.3 \times
10^{16}\ \mdotrxj$, much lower than the secular value of $1.9 \times 10^{18}\ \mbox{g s}^{-1}$ predicted for
its long 7.2 hr orbital period. From the detection of circular polarization in the optical range, that peaks
in the $V$ band, \citet{katajainen07} suggest a magnetic moment $\mu$ in the range $3 \times 10^{33} - 3 \times
10^{34}$. The condition for accretion would then imply $\dot{M} \gtrsim 5.3 \times 10^{17}\ \mbox{g s}^{-1}$.
Although this value is probably at the high side, it might indicate that the a substantial contribution to the
accretion luminosity comes from cyclotron radiation and not from the hard X-rays.

Because of the wide range of temperatures found for the soft X-ray component, \citet{anzolin08} proposed that
$T_{\mathrm{BB}}$ could be related to the magnetic field strength. Lower temperatures would indicate higher
field systems because of larger irradiated WD areas by cyclotron radiation. However, the recent detection of
significant polarized emission in RXJ2133 \citep{katajainen07} and 1RXS~J173021.5-055933  \citep{butters09} adds
them to the subset of soft X-ray and polarized IPs together with V2400 Oph, PQ Gem and V405 Aur. While the
degree of polarization could also be affected by the accretion geometry, these five polarized systems possess
soft blackbody components that span a wide range of temperatures, with three out of five displaying hot blackbodies.
The inferred magnetic field strengths are quite uncertain for most systems, being 9--21 MG for PQ Gem
\citep{potter97}, 9--27 MG for V2400 Oph \citep{vaeth97} and $\sim 30$ MG for V405 Aur \citep{piirola08}.
A 20 MG field was proposed for RXJ2133 \citep{katajainen07}, while no field estimate is given for
1RXS~J173021.5-055933 \citep{butters09}. To really test the proposal made by \citet{anzolin08},
spectro-polarimetric measures of the magnetic field strengths are essential for all soft systems.
Furthermore, there are indications that the reprocessed radiation emerges in the UV range also in IPs, as 
found in PQ Gem \citep{stavroyiannopoulos97} and UU Col \citep{demartino06a}. This aspect is essential to
properly determine the reprocessed energy budget.

\section{Conclusions}

The two CVs IGR0023 and RXJ2133 have been confirmed as true members of the IP class by using pointed X-ray
observations with \emph{XMM-Newton} and \emph{Suzaku} satellites. A strong pulsation at the WD rotational
period dominates the power spectra of IGR0023 below 2 keV, while for RXJ2133 it is detected up to 12 keV.
Both systems have fast rotating WDs, with $P_{\omega}$ of 561 s for IGR0023 and 571 s for RXJ2133. The latter
is one of the most asynchronous systems among IPs, having a ratio between the spin and the orbital periods of
0.022. The fast rotation and very long orbital period, together with the detection of substantial polarized 
emission, suggests that RXJ2133 is likely a young IP that will evolve into a polar when attaining synchronism.

Their broad-band X-ray spectra were analyzed including also \emph{INTEGRAL} data, that allowed us to cover the
energy range 0.2--100 keV. These are well described by a multi-temperature plasma emission, with a minimum
temperature of 0.2--0.3 keV and a maximum temperature of $\sim 50$ keV, which implies relatively massive WDs.
While this might suggest that hard X-ray CVs harbor massive primaries, it has still to be understood
if this is the only ingredient for a CV to be a hard X-ray source.

In RXJ2133, a $\sim 100$ eV blackbody emission is also required to fit the soft portion of the spectrum.
The temperature of this component is similar to that found in V2400 Oph \citep{demartino04} and
1RXS~J173021.5-055933 \citep{demartino08}, both of them also showing circularly polarized radiation.
It is however different from that found in the other two soft X-ray and polarized IPs, PQ Gem and V405 Aur 
\citep{demartino04,anzolin08}. This casts doubts on the possible relation between the soft X-ray temperature
and the magnetic field strength proposed by \citet{anzolin08}. The present work also opens a further question
on whether the reprocessed radiation in the soft and polarized IPs also emerges at UV wavelengths, as recently
demonstrated by \citet{konig06} for the polar prototype AM Her.

Furthermore, the fact that RXJ2133, V2400 Oph and 1RXS~J173021.5-055933 are bright hard X-ray sources and
are also polarized IPs might not imply that cyclotron cooling decreases the hard X-ray flux
\citep{woelk_beuermann96,fischer_beuermann01},
at least for the magnetic strength values covered by these IPs. Observations aiming at measuring the magnetic
field strength of IPs will help in making a clear picture of emission mechanisms in these systems.

\begin{acknowledgements}
We acknowledge the \emph{XMM-Newton} MSSL and SOC staff for help in the reduction of the OM data. DdM and GA
acknowledge financial support from ASI under contract ASI/INAF I/023/05/06 and ASI/INAF I/088/06/0 and
from INAF under contract PRIN-INAF 2007 N.17. We also acknoweledge useful suggestions and comments from the
referee, prof. Klaus Beuermann, that helped us to improve this work.
\end{acknowledgements}

\bibliographystyle{aa}
\bibliography{0000}

\begin{landscape}

\begin{table}
\begin{minipage}{23cm}
\caption{Spectral parameters of the best-fit models to the EPIC-pn, 
combined MOS and IBIS/ISGRI
average spectra of IGR0023.The different models (see text) are indicated with A, B and C in the first column.
Errors indicate the $90 \%$ confidence level of the corresponding parameter.}
\label{tab:spectra0023}
\centering
\renewcommand{\footnoterule}{}
\begin{tabular}{llllllllllllll}
\hline\hline       
& $N_{\mathrm{H}}$ \footnote{Column density of the total absorber in units of
$10^{21}\ \mathrm{cm}^{-2}$.} &
$N_{\mathrm{H}}$ \footnote{Column density of the partial absorber in units of
$10^{23}\ \mathrm{cm}^{-2}$.} &
$C_F$ \footnote{Covering fraction of the partial absorber.} &
$A_Z$ \footnote{Metal abundance.} &
$\alpha$ \footnote{Index of the power-law emissivity function of the CEMEKL component.} &
$k T_1$ &
$C_1$ \footnote{Normalization constant of the first optically thin plasma component in units
of $10^{-4}$.} &
$k T_2$ &
$C_2$ \footnote{Normalization constant of the second optically thin plasma component in units
of $10^{-3}$.} &
$k T_3$ \footnote{In model A this the maximum temperature of the CEMEKL.} &
$C_3$ \footnote{Normalization constant of the third optically thin plasma component (CEMEKL
in model A) in units of $10^{-2}$.} &
E.W. \footnote{Equivalent width of the 6.4 keV Fe line.} &
$\chi_{\nu}^2$ \\
 & & & & & & (keV) & & (keV) & & (kev) & & (eV) & ($\chi^2$ / d.o.f.) \\
\hline
A & $1.79 \pm 0.07$ & $0.8 \pm 0.2$ & $0.39_{-0.06}^{+0.05}$ & $0.7 \pm 0.2$ & $1.8_{-0.4}^{+0.6}$
& & & & & $28_{-11}^{+19}$ & $1.7_{-0.5}^{+1.3}$ & $100_{-25}^{+30}$ & 0.86 (892/1033)\\
B & $2.0 \pm 0.1$ & $1.1_{-0.2}^{+0.3}$ & $0.44 \pm 0.04$ & $0.5 \pm 0.1$ & & $0.17_{-0.04}^{+0.02}$ &
$3.0_{-1.7}^{+3.3}$ & $14 \pm 2$ & $5.2_{-0.3}^{+0.2}$ & & & $100_{-29}^{+24}$ & 0.86 (886/1032) \\
C & $2.0 \pm 0.1$ & $0.9_{-0.2}^{+0.3}$ & $0.36 \pm 0.06$ & $0.7_{-0.2}^{+0.3}$ & &
$0.17_{-0.05}^{+0.03}$ & $1.5_{-1.0}^{+2.3}$ & $9_{-3}^{+2}$ & $2.1_{-0.9}^{+1.5}$ &
$> 27$ & $0.3 \pm 0.1$ & $94_{-27}^{+24}$ & 0.85 (870/1030) \\
\hline
\end{tabular}
\end{minipage}
\end{table}

\begin{table}
\begin{minipage}{23cm}
\caption{Spectral parameters of the best-fit models to the EPIC-pn, 
combined MOS and IBIS/ISGRI
average spectra of RXJ2133 for May 2005 (\textit{1}) and July 2005 (\textit{2}), as well as to the
XIS and HXD spectra (\textit{3}). The different models (see text) are indicated with A, B and C in the first column.
Errors indicate the $90 \%$ confidence level of the corresponding parameter.}
\label{tab:spectra2133} 
\centering
\renewcommand{\footnoterule}{}
\begin{tabular}{lllllllllllllllll}
\hline\hline       
& &
$N_{\mathrm{H}}$ \footnote{Column density of the total absorber in units of
$10^{21}\ \mathrm{cm}^{-2}$.} &
$N_{\mathrm{H}}$ \footnote{Column density of the partial absorber in units of
$10^{23}\ \mathrm{cm}^{-2}$.} &
$C_F$ \footnote{Covering fraction of the partial absorber.} &
$k T_{\mathrm{BB}}$ &
$C_{\mathrm{BB}}$ \footnote{Normalization constant of the BBODY component in units of $10^{-4}$.} &
$A_Z$ \footnote{Metal abundance.} &
$\alpha$ \footnote{Index of power-law emissivity function of the CEMEKL component.} &
$k T_1$ &
$C_1$ \footnote{Normalization constant of the first optically thin plasma component in units
of $10^{-4}$.} &
$k T_2$ &
$C_2$ \footnote{Normalization constant of the second optically thin plasma component in units
of $10^{-3}$.} &
$k T_3$ \footnote{In model A this the maximum temperature of the CEMEKL.} &
$C_3$ \footnote{Normalization constant of the third optically thin plasma component (CEMEKL
in model A) in units of $10^{-2}$.} &
E.W. \footnote{Equivalent width of the 6.4 keV Fe line.} &
$\chi_{\nu}^2$ \\
 & & & & & (eV) & & & & (keV) & & (keV) & & (keV) & & (eV) & ($\chi^2$ / d.o.f.) \\
\hline
& \textit{1} & $1.7_{-0.2}^{+0.1}$ & $1.3_{-0.1}^{+0.2}$ & $0.56_{-0.04}^{+0.03}$ & $94_{-2}^{+3}$ &
$5.8_{-1.1}^{+0.8}$ & $0.6_{-0.1}^{+0.2}$ & $0.9_{-0.1}^{+0.2}$ & & & & & $55_{-15}^{+21}$ &
$3.5 \pm 0.4$ & $159 \pm 22$ & 0.98 (1416/1444) \\
A & \textit{2} & 1.7 (fixed) & $1.1 \pm 0.2$ & $0.58_{-0.04}^{+0.03}$ & $97_{-1}^{+2}$ &
$4.9_{-0.5}^{+0.4}$ & $0.6_{-0.1}^{+0.2}$ & $0.9 \pm 0.2$ & & & & & $45_{-10}^{+23}$ & $3.5 \pm 0.6$ &
$149_{-31}^{+16}$ & 1.05 (819/780) \\
& \textit{3} & 1.7 (fixed) & $1.06 \pm 0.07$ & $0.48_{-0.02}^{+0.01}$ & $102 \pm 1$ & $3.7 \pm 0.2$ &
$0.8 \pm 0.1$ & 1 (fixed) & & & & & $82_{-10}^{+13}$ & $3.80_{-0.08}^{+0.09}$ & $181 \pm 10$ &
1.12 (2907/2608) \\
\hline
& \textit{1} & $1.7_{-0.08}^{+0.12}$ & $1.4 \pm 0.2$ & $0.52_{-0.02}^{+0.04}$ & $98 \pm 2$ &
$4.9_{-0.6}^{+1.0}$ & $0.5 \pm 0.1$ & & & & $1.4_{-0.2}^{+0.4}$ & $0.9_{-0.3}^{+0.5}$ &
$27_{-6}^{+4}$ & $1.70_{-0.07}^{+0.05}$ & $168_{-24}^{+23}$ & 0.99 (1431/1443) \\
B & \textit{2} & 1.7 (fixed) & $1.2 \pm 0.2$ & $0.59 \pm 0.04$ & $103 \pm 2$ &
$4.3_{-0.4}^{+0.5}$ & $0.7_{-0.2}^{+0.3}$ & & & & $4.6_{-1.0}^{+1.9}$ & $5.1_{-1.7}^{+1.9}$ &
$36_{-8}^{+13}$ & $1.3 \pm 0.2$ & $140 \pm 30$ & 1.06 (822/779) \\
& \textit{3} & 1.7 (fixed) & $1.03 \pm 0.07$ & $0.46_{-0.01}^{+0.02}$ & $107 \pm 1$ &
$3.1 \pm 0.1$ & $0.9_{-0.1}^{+0.2}$ & & & & $7.3_{-1.1}^{+0.6}$ & $3.6_{-0.9}^{+0.8}$ &
$> 71$ & $1.46_{-0.06}^{+0.07}$ & $173 \pm 9$ & 1.08 (2818/2606) \\
\hline
& \textit{1} & $1.7 \pm 0.1$ & $1.3 \pm 0.2$ & $0.54 \pm 0.03$ & $96_{-2}^{+3}$ &
$5.4_{-0.9}^{+0.8}$ & $0.7 \pm 0.2$ & & $1.1_{-0.2}^{+0.3}$ & $4.1_{-1.7}^{+3.3}$ &
$4.8_{-0.9}^{+3.2}$ & $3.3_{-0.5}^{+1.5}$ & $39_{-9}^{+11}$ & $1.46_{-0.18}^{+0.09}$ &
$156_{-21}^{+22}$ & 0.97 (1403/1441) \\
C & \textit{2} & 1.7 (fixed) & $1.1 \pm 0.2$ & $0.57 \pm 0.04$ & $99 \pm 2$ &
$4.7 \pm 0.5$ & $0.7_{-0.2}^{+0.3}$ & & $1.0_{-0.2}^{+0.3}$ & $4.5_{-2.3}^{+3.6}$ &
$5.4_{1.4}^{+1.7}$ & $4.7_{-1.7}^{+2.4}$ & $39_{-8}^{+12}$ & $1.3_{-0.2}^{+0.1}$ &
$142_{-30}^{+31}$ & 1.04 (805/777) \\
& \textit{3} & 1.7 (fixed) & $1.10 \pm 0.09$ & $0.45_{-0.02}^{+0.01}$ & $100_{-3}^{+2}$ &
$4.7_{-0.8}^{+1.0}$ & $0.9_{-0.1}^{+0.2}$ & & $0.9_{-0.1}^{+0.2}$ & $1.6_{-0.6}^{+0.9}$ &
$6.6 \pm 1.0$ & $3.5_{-0.9}^{+1.0}$ & $> 61$ & $1.50_{-0.05}^{+0.07}$ & $170 \pm 9$ &
1.07 (2794/2603) \\
\hline
\end{tabular}
\end{minipage}
\end{table}

\end{landscape}

\end{document}